\documentclass[runningheads]{llncs}
\usepackage{amsmath}
\usepackage{booktabs} 
\usepackage{caption} 
\usepackage{graphicx}
\usepackage{pgfplots}
\usepackage[all]{nowidow}
\usepackage[utf8]{inputenc}
\usepackage{tikz}
\usetikzlibrary{er,positioning,bayesnet}
\usepackage{multicol}
\usepackage{hyperref}
\usepackage[inline]{enumitem} 
\definecolor{blue}{HTML}{1F77B4}
\definecolor{orange}{HTML}{FF7F0E}
\definecolor{green}{HTML}{2CA02C}

\pgfplotsset{compat=1.14}

\usepackage{multirow}
\usepackage{amsmath,amsfonts,mathtools}
\usepackage{diagbox}
\usepackage{amssymb}

\usepackage{algorithm}
\usepackage[noend]{algpseudocode}

\algnewcommand\algorithmicforeach{\textbf{for each}}
\algdef{S}[FOR]{ForEach}[1]{\algorithmicforeach\ #1\ \algorithmicdo}

\usepackage[font=footnotesize]{caption}
\usepackage{subfig}
\usepackage{booktabs} 
\usepackage{multirow}
\usepackage{amsmath,amsfonts,mathtools}
\usepackage{diagbox}

\algtext*{EndWhile}
\algtext*{EndIf}
\algtext*{EndFor}

\usepackage{pifont}
\newcommand{\cmark}{\ding{51}}%
\newcommand{\xmark}{\ding{55}}%

\begin{document}

\title{Threats on Logic Locking: A Decade Later}

\author{Kimia Zamiri Azar \and Hadi Mardani Kamali \and \\ Houman Homayoun \and Avesta Sasan}

\institute{Department of Electrical and Computer Engineering, \\
George Mason University, Fairfax, VA, USA. \\
\email{\{kzamiria, hmardani, hhomayou, asasan\}@gmu.edu}}

\maketitle

\begin{abstract}

To reduce the cost of ICs and to meet the market's demand, a considerable portion of manufacturing supply chain, including silicon fabrication, packaging and testing may be pushed offshore. Utilizing a global IC manufacturing supply chain, and inclusion of non-trusted parties in the supply chain has raised concerns over security and trust related challenges including those of overproduction, counterfeiting, IP piracy, and Hardware Trojans to name a few. To reduce the risk of IC manufacturing in an untrusted and globally distributed supply chain, the researchers have proposed various locking and obfuscation mechanisms for hiding the functionality of the ICs during the manufacturing, that requires the activation of the IP after fabrication using the key value(s) that is only known to the IP/IC owner. At the same time, many such proposed obfuscation and locking mechanisms are broken with attacks that exploit the inherent vulnerabilities in such solutions. The past decade of research in this area, has resulted in many such defense and attack solutions. In this paper, we review a decade of research on hardware obfuscation from an attacker perspective, elaborate on attack and defense lessons learned, and discuss future directions that could be exploited for building stronger defenses. 
 
\end{abstract}


\section{Introduction}

The increasing cost of IC manufacturing has pushed several stages of the semiconductor device's manufacturing supply chain offshore \cite{yeh2012trends}. However, many of these offshore facilities are identified as untrusted entities. Processing and fabrication of ICs in an untrusted supply chain poses a number of challenging security threats such as IP piracy and IC overproduction \cite{rostami2014primer}. To counter these threats, various hardware design-for-trust techniques have been proposed. The term \emph{logic locking}, \emph{a.k.a.} hardware obfuscation, surfaced in 2008 by EPIC \cite{roy2010ending}, in which a limited programmability was introduced into a netlist by means of inserting additional key programmable gates (KG)s at design time. After fabrication, the functionality of the IC is programmed by loading the correct key values. The key inputs could be stored in, and loaded from, an on-chip tamper-proof memory \cite{tuyls2006read}. The purpose of inserting KGs is to hide the IC's functionality from untrusted foundries. Since the functionality of a design is locked with a secret key, the attacker cannot learn the functionality of the locked netlist after reverse engineering. 
Insertion of $n$ KGs hides the ICs functionality between $2^n$ different circuit possibilities, each generated by a different key. The correct functionality will be recovered when the loaded $n$-bit key is correct.  

\begin{figure}[t]
\centering 
\includegraphics[width = 350pt]{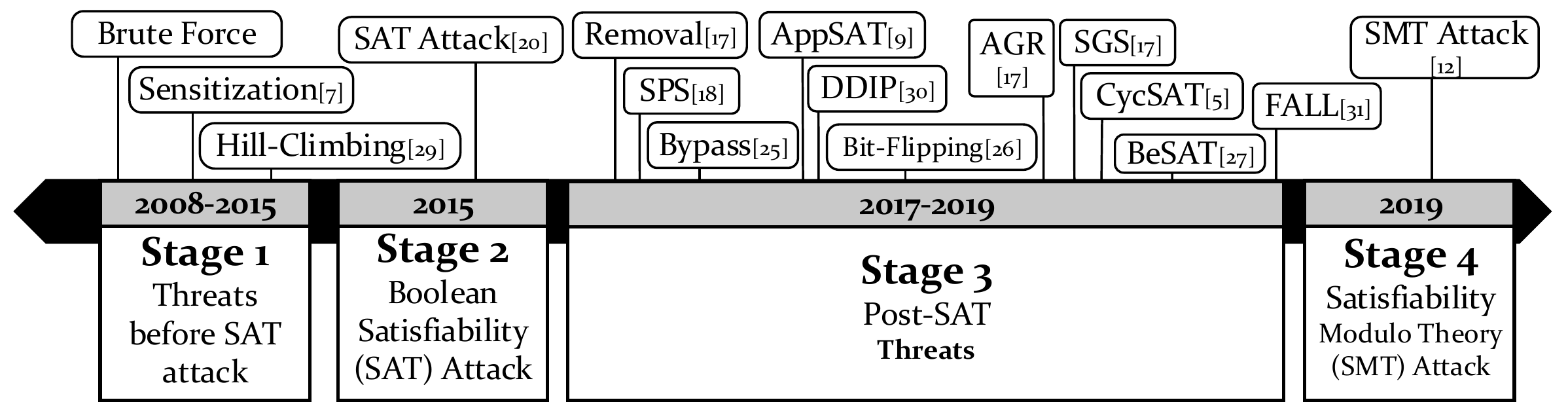}
\caption{Categorization of attacks against logic locking schemes.}
\label{Attacks}
\end{figure}

EPIC, however did not end the threat against IP piracy (or other related concerns), as this solution and many other obfuscation solutions that were proposed over the last decade were broken using various carefully crafted attacks. A decade of research in this area, has resulted in a wide range of defense \cite{baumgarten2010preventing}, \cite{kamali2018lut}, \cite{kamali2019full}, \cite{rajendran2012security}, \cite{rajendran2015fault}, \cite{shamsi2017cyclic}, \cite{li2017provably}, \cite{yasin2016sarlock}, \cite{yasin2017provably}, \cite{yasin2017lock}, \cite{SRCLock}, \cite{winograd2016hybrid}, \cite{xie2017delay}, \cite{xie2016antisat} and attack solutions \cite{chakraborty2018timingsat}, \cite{zhou2017cycsat}, \cite{rajendran2012security}, \cite{shamsi2017appsat}, \cite{azar2019smt}, \cite{yasin2017removal}, \cite{Yasin2017sps}, \cite{subramanyan2015evaluating}, \cite{roshanisefat2018benchmarking}, \cite{xu2017novel}, \cite{shen2018sat}, \cite{shen2019besat}, \cite{plaza2015solving}, \cite{shen2017double}, \cite{subramanyan1functional}. In this paper, we review many of these obfuscation solutions, explain and reviewing most notable attack mechanisms, summarize and compare the effectiveness of obfuscation solutions against these attacks, and describe the strength and weaknesses of various obfuscation and attack solutions. As illustrated in Fig. \ref{Attacks}, the defense and attack solutions related to hardware obfuscation, based on functionality, capability, effectiveness and time-line are categorized into four categories: (1) \emph{Test-Inspired Attacks} that were mostly inspired from test concepts and attempted to discover the obfuscation key value based on the location of KGs, described in Section \ref{stage1}. (2) SAT Attack, formulation and revelation of which significantly affected the direction and presumed assumptions of the hardware obfuscation research community, explained in Section \ref{stage2}. (3) Post-SAT Attacks where the focus of hardware security researchers changed to the design of an attack against obfuscation solutions that resist the SAT attack, explained in \ref{stage3}. And (4) SMT Attack as a universal attack platform capable of instantiating many theory solvers to act as pre- post- or co- processors to the SAT solver, described in Section \ref{stage4}. We conclude the paper in Section \ref{stage5} by summarizing the effectiveness of attacks discussed in this paper and provide a short discussion on new opportunities related to designing secure logic locking solutions. 

\section{Stage 1: Test-Based Attacks} \label{stage1}

\subsection{Brute Force Attack}
The \emph{brute force} attack is the most intuitive attack against obfuscated circuits. This attack exhaustively search for the correct key by testing all key and input values.
For instance, assuming that adversary has access to the reverse-engineered netlist, and considering that the circuit has four $PI$s ( $i_{0..4}$) and two $KI$s ($k_{0..1}$), an exhaustive search may result in applying of $2^{2+4} = 64$ test patterns (in the worst case) and checking the output against an activated (functionally correct) chip to verify correctness. Based on the number of primary inputs ($|PI|$) and the number of key bits ($|KI|$), the number of possible test patterns is ($2^{|PI|+|KI|}$). Hence, the search space for a brute force attack is extremely large, making the attack even for small circuits and small number of keys unfeasible in a reasonable amount of time. For example, a small circuit with 20 input pins, which is obfuscated with 80 key gates poses $2^{100}$ possible test pattern. An attacker can reduce the number of test patterns using functional test or random test, in which the exponential impact of $|PI|$s will be eliminated, and only $2^{|KI|}\times(func\_test\_patterns)$ is required for brute force attack. But even with this change, the attack time is exponential with respect to the number of key gates. EPIC \cite{roy2010ending} used a random KG insertion policy referred to as random logic locking (RLL). Using RLL, EPIC reasoned that by replacing a small percentage of gates (or insertion of KGs), the obfuscation can resist brute force attacks.


\subsection{Sensitization Attack}

After introducing EPIC \cite{roy2010ending}, Rajendran \emph{et al.} \cite{rajendran2012security} proposed a \emph{sensitization} attack, which determines individual key values, in a time linear to the |KI|, by applying patterns that sensitize key values to the output. As its name implies, sensitization of an internal wire (key bit) $\bf L$ to an output $\bf O$ means that the value of $\bf L$ can be propagated to $\bf O$ and thus any change on $\bf L$ is observable on $\bf O$. After determining an input pattern that propagates the value of the key-bit to the output, the attacker applies the input pattern to a functional IC (An IC activated and programmed with the correct key that could be obtained from market). The correct key value will be propagated to an output by applying this pattern to the functional IC. The attacker observe and record this output as the value of the sensitized key-bit. The propagation of a key-bit to the output is heavily depending on the location of the KGs, hence, they classify KGs based on their location and discuss corresponding attack strategies for each case. The summary of strategies and techniques used in the sensitization attack is reflected in Table \ref{Sensitization}. To prevent sensitization attack they proposed SLL, in which the KGs are inserted in locations with maximum mutual interference. In SLL the attacker cannot sensitize the key-bit values to a primary output. Similar to SLL, several prior-art methods in the literature, including fault-analysis (FLL), LUT-based locking, etc. \cite{baumgarten2010preventing}, \cite{kamali2018lut}, \cite{rajendran2012security}, \cite{rajendran2015fault}, \cite{winograd2016hybrid}, , tried to maximize the complexity of obfuscation using different KGs replacement strategies.

\begin{table}[t]
\scriptsize
\centering
\caption{Classification of KGs in Sensitization Attack.}
\label{Sensitization}
\setlength\tabcolsep{10pt} 
\begin{tabular}{@{} l l l @{}}
\toprule  
\textbf{Term}    & \textbf{Description} & \textbf{Strategy used by attacker}  \\ \hline
\textbf{Runs of KGs}     & Back-to-Back KGs &  Replacing by a Single KG\\ \hline
\multirow{2}{*}{\textbf{Isolated KGs}}     & \multirow{2}{*}{No Path between KGs} & Finding Unique Pattern per\\
& & KG (Golden Pattern (GP)) \\ \hline
\multirow{2}{*}{\textbf{Dominating KGs}}     &  $k1$ is on Every Path & Muting $k0$, \\
& between $k0$ and $PO$s & Sensitizing $k1$ \\ \hline
\textbf{Concurrently Mutable}     &  Convergent at a Third Gate & Muting $k0/k1$, \\
\textbf{Convergent KGs} & Both can be Propagated to $PO$s &  Sensitizing $k1/k0$ \\ \hline
\textbf{Sequentially Mutable}     & Convergent at a Third Gate  & Determining $k1$ by GP, \\ 
\textbf{Convergent KGs} & One can be Propagated to $PO$s & Update the Netlist, Target $k0$     \\ \hline

\textbf{Non-Mutable}     & Convergent at a Third Gate  & \multirow{2}{*}{Brute Force Attack} \\ 
\textbf{Convergent KGs} & None can be Propagated to $PO$s &  \\\bottomrule

\end{tabular}
\end{table}

\subsection{Random-based Hill-Climbing Attack}

Plaza \emph{et al.} \cite{plaza2015solving} developed a new algorithmic attack that uses test patterns and observe responses. Unlike sensitization attack \cite{rajendran2012security}, their proposed approach does not require netlist access. They propose a randomized local key-searching algorithm to search the key that can satisfy a subset of correct input/output patterns. The algorithm proposed in \cite{plaza2015solving} is iterative in nature. At first, it selects random value for key bits and then at each iteration, the key bits, which are selected randomly, are toggled one by one. The target is to minimize the frequency of differences between the observed and expected responses. Hence, a random key candidate is gradually improved based on observed test responses.  If no solution is found in one iteration, the algorithm resets the key to a new random key value. However, the complexity of this attack quickly increases with increasing number of KGs.

\section{Stage 2: SAT Attack} \label{stage2}

\begin{figure}[t]
\centering
\includegraphics[width = 350pt]{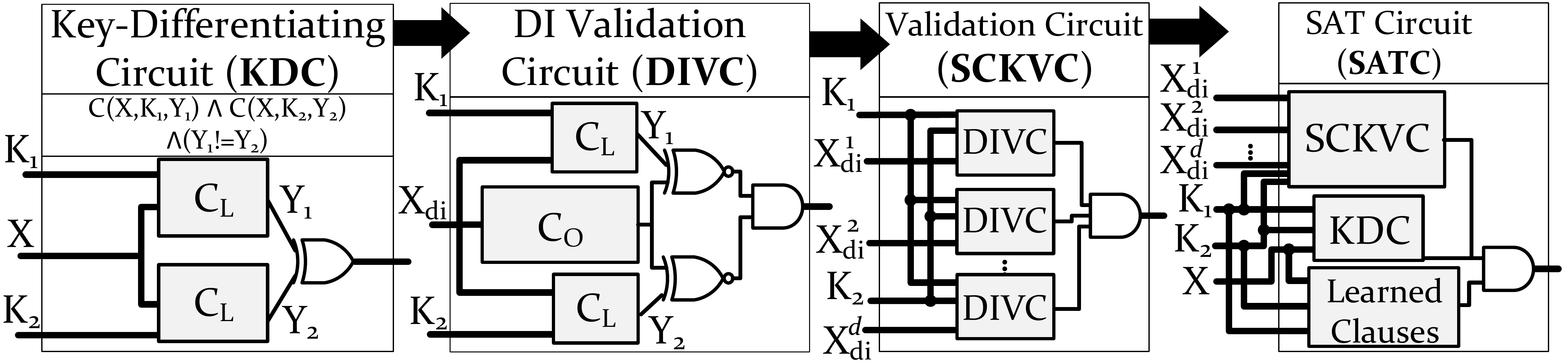}
\caption{SAT Attack Iterative Flow.}
\label{satc}
\end{figure}

In 2015, Subramanyan \emph{et al.} \cite{subramanyan2015evaluating} proposed a new and powerful attack using Boolean satisfiability (SAT) solver, called SAT attack, that effectively and quickly broke all previously proposed logic locking techniques. As an "\emph{oracle-guided}" attack, SAT attack requires a reverse-engineered but locked netlist ($C_L$), and a functionally activated chip ($C_O$). A circuit view of steps taken in a SAT attack is shown in Fig. \ref{satc}. For this attack, the attacker first replicate the obfuscated circuit and builds a double circuit which is used for finding an input (X$_d$[i]) that for two different key values generates two different outputs. Such input is referred to as Discriminating Input Pattern(DIP).  Each X$_d$[i] is used to create a DI validation circuit (DIVC). The validation circuit, as shown in Fig. \ref{satc} assures that for a previously found X$_d$[i], two different keys generate the same output value. Each iteration of the SAT attack finds a new (X$_d$[i]), and add a new DI validation circuit. The DIVCs are ANDed together to form a Set of Correct Key Validation Circuit (SCKVC). In each iteration, the SAT solver try to find a new X$_d$[i] and two key values that satisfy the double circuit (KDC) and the Validation Circuit (SCKVC). The key values and the X$_d$[i], as illustrated in Alg. \ref{SAT} (line 5), is found by a SAT query. This means the new key generate two different values for the new X$_d$[i], but generate the same value for all previously found X$_d$s for both key values. This process continues until the SAT solver cannot find a new X$_d$[i] (line 4). At this point any key that generates the correct output for the set of found X$_d$s is the correct key (line 9).


\begin{algorithm}
\caption{SAT-based Attack Algorithm \cite{subramanyan2015evaluating}}
\label{SAT}
\begin{algorithmic}[1]
\scriptsize
\Function{SAT\_Attack}{Circuit~C$_{L}$, Circuit C$_{O}$}
    \State \emph{i} $\gets$ 0; \emph{F$_{0}$} $\gets$ C$_{L}$(X, K$_{1}$, Y$_{1}$) $\land$ C$_{L}$(X, K$_{2}$, Y$_{2}$);
    \While {\emph{SAT}(\emph{F$_{i}$} $\land$ (Y$_{1}$ $\neq$ Y$_{2}$))} 
        \State X$_d$[i] $\gets$ sat\_assignment (F$_{i} \land $(Y$_1 \neq $Y$_2$)); Y$_d$[i] $\gets$ C$_{O}$(X$_d$[i]);
        \State \emph{F$_{i+1}$} $\gets$ \emph{F$_{i}$} $\land$ C$_{L}$(X$_d$[i], K$_{1}$, Y$_d$[i]) $\land$ C$_{L}$(X$_d$[i], K$_{2}$, Y$_d$[i]); \emph{i} $\gets$ \emph{i+1};
    \EndWhile
    \State \emph{$K^*$}  $\gets$ sat\_assignment$_{K_1}$(\emph{F$_{i}$});
\EndFunction
\end{algorithmic}
\end{algorithm}

\begin{figure}[t]
\begin{minipage}{0.33\linewidth}
\centering
\includegraphics[width=\linewidth]{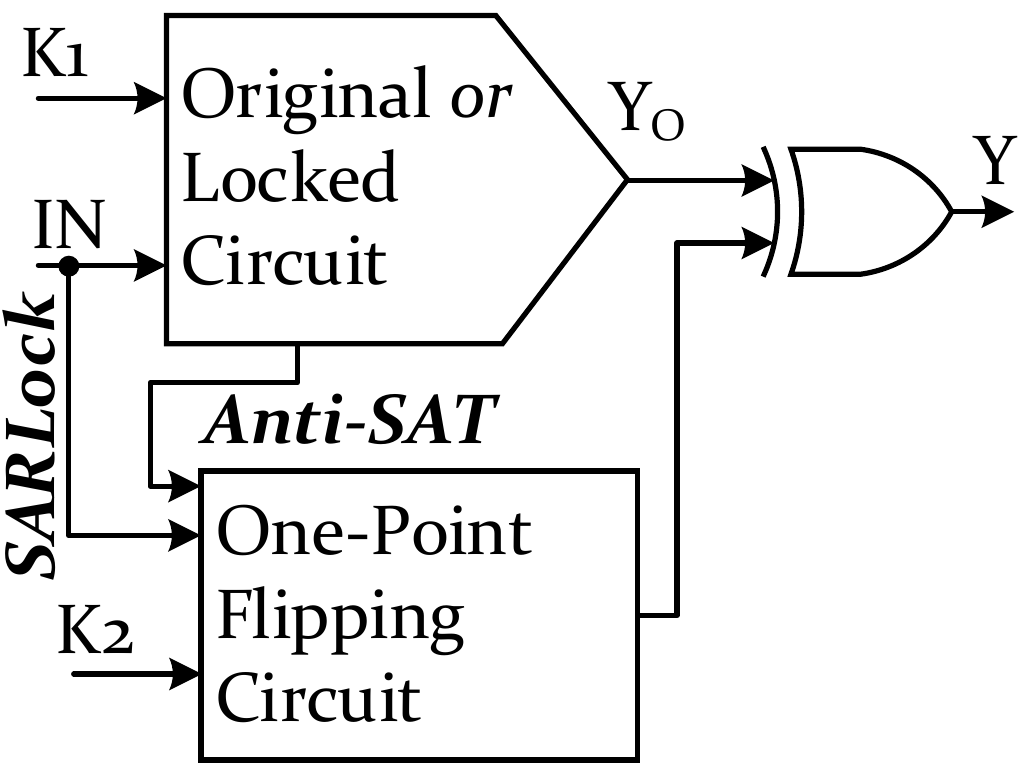}
\end{minipage}
\hspace{35pt}
\begin{minipage}{0.47\linewidth}
\scriptsize
\centering
\setlength\tabcolsep{2.5pt} 
\begin{tabular}{@{} l *{10}c @{}}
\toprule
$Y_O$ & IN & k=0 & k=1 & k=2 & k=3 & k=4 & k=5 & k=6 & k=7 \\
\midrule
   \cmark    &  0 & \cmark & \cmark & \cmark & \cmark & \textcolor{red}{\xmark} & \cmark & \cmark & \cmark \\
   \cmark    &  1 & \textcolor{red}{\xmark} & \cmark & \cmark & \cmark & \cmark & \cmark & \cmark & \cmark \\
   \cmark    &  2 & \cmark & \cmark & \textcolor{red}{\xmark} & \cmark & \cmark & \cmark & \cmark & \cmark \\
   \cmark    &  3 & \cmark & \cmark & \cmark & \cmark & \cmark & \textcolor{red}{\xmark} & \cmark & \cmark \\
   \cmark    &  4 & \cmark & \cmark & \cmark & \cmark & \cmark & \cmark & \cmark & \cmark \\
   \cmark    &  5 & \cmark & \cmark & \cmark & \cmark & \cmark & \cmark & \textcolor{red}{\xmark} & \cmark \\
   \cmark    &  6 & \cmark & \cmark & \cmark & \cmark & \cmark & \cmark & \cmark & \textcolor{red}{\xmark}\\
   \cmark    &  7 & \cmark & \cmark & \cmark & \textcolor{red}{\xmark} & \cmark & \cmark & \cmark & \cmark \\ \bottomrule
\end{tabular}
\end{minipage}\hfill
\caption{Flipping Structure of SARLock and Anti-SAT.}
\label{sarlock_anti}
\end{figure}

For all prior obfuscation schemes, even those resistant to sensitization attack, the SAT attack was able to rule out a significant number of key values at each iterations (by finding each DIP). Hence, In order to thwart SAT attack, the first attempted approach was to weaken the strength of the DIPs to reduce its pruning power. SARLock \cite{yasin2016sarlock} and Anti-SAT \cite{xie2016antisat} were the first prior-art methods that accomplished this. Both SARLock and Anti-SAT engaged one-point flipping function, demonstrated in Fig. \ref{sarlock_anti}. Using this obfuscation scheme, each DIP is able to rule out only one incorrect key. Hence, the SAT attack requires to apply all $2^{|KI|}$ to retrieve the correct functionality. However, this method results in obfuscation circuits that for all but one output work as the original circuit, and the output corruption upon application of a wrong key is quite low.

\section{Stage 3: Post-SAT Attacks} \label{stage3}

As discussed, the proposed SAT-resilient solutions suffered from low output corruption. This however could have been addressed by combining a SAT-hard solution with a traditional obfuscation solution, such as RLL or SLL, that exhibits high level of output corruption. Although SAT-resilient logic locking schemes provided a defense against SAT attack, researchers found new vulnerabilities associated with this class of obfuscation techniques resulting in the development of many new attacks on the presumed SAT-hard logic locking schemes described in this section.

\subsection{Removal Attack}

As shown in Fig \ref{sarlock_anti}, in bare implementation of one-point flipping circuit, the locking circuitry is completely decoupled from the original circuit. A removal attack identifies and removes/bypasses the locking circuitry to retrieve the original circuit and to remove dependence on key values \cite{yasin2017removal}. The removal attack, presented in \cite{yasin2017removal}, was used to detect and remove SARLock \cite{yasin2016sarlock}. In presence of removal attack, researchers investigated SAT-hard solutions that are hard to detect (preventing removal by pure structural analysis), an example of which was Anti-SAT \cite{xie2016antisat}.

\subsection{Signal Probability Skew (SPS) Attack}
The Signal Probability Skew (SPS) attack \cite{Yasin2017sps} leverages the structural traces in Anti-SAT block to identify and isolate the Anti-SAT block \cite{xie2016antisat}. \emph{Signal probability skew} (SPS) of a signal \emph{x} is defined as $s=P_r[x=1]-0.5$, where $P_r[x=1]$ indicates the probability that signal $x$ is $1$. The range of $s$ is $[-0.5,0.5]$. If the SPS of signal $x$ is closer to zero, an attacker have lower chance of guessing the signal value by random. For a 2-input gate, the signal probability skew is the difference between the signal probability of its input wires. The flipping-circuit in the Anti-SAT is constructed using two complementary circuits, $g$ and $\overline{g}$, in which the number of input vectors that make the function $g$ equal to 1 ($p$) is either close to 1 or $2^{n}-1$. These two complementary circuits converge at an AND gate $G$. Considering this structure, the \emph{absolute difference of the signal probability skew} (ADS) of the inputs of gate $G$ is quite large, noting that the SAT resilience is ensured by this skewed $p$. Algorithm \ref{SPS} shows the SPS attack, which identifies the Anti-SAT block's output by computing signal probabilities and searching for the skew(s) of arriving signals to a gate in a given netlist.



\begin{algorithm}
\caption{SPS Attack Algorithm \cite{Yasin2017sps}}
\label{SPS}
\begin{algorithmic}[1]
\scriptsize
\Function{SPS\_Attack}{Circuit~C$_{L}$}
    \State \emph{ADS}$_{arr}$ $\gets$ $\{\}$;
    \ForEach{\emph{gate} $\in$ C$_{L}$}
        \State \emph{ADS}$_{arr}$(gate$_{i}$) $\gets$ Compute\_ADS(C$_{L}$, gate$_{i}$);
    \EndFor
    \State $G$ $\gets$ Find\_Maximum(\emph{ADS}$_{arr}$);
    \State $Y$ $\gets$ Find\_value\_from\_skew($G$); \Comment{Correct value of Anti\_SAT output}
    \State C$_{Lock}$ $\gets$ remove\_\textbf{TFI}(C$_{L}$, $G$, $Y$); \Comment{Transitive FanIn of the gate $G$}
    \State \Return C$_{Lock}$ \Comment{C$_{Lock}$: C$_{L}$ after removing Anti\_SAT block}
\EndFunction
\end{algorithmic}
\end{algorithm}

\subsection{Bypass Attack}

Although SARLock and Anti-SAT break the SAT attack, their respective output corruptibility is very low if they are not mixed with traditional logic locking, such as SLL. Observing and relying on the very low level of output corruption in such SAT-hard solutions, the bypass attack  \cite{xu2017novel} was introduced. The bypass attack instantiates two copies of the obfuscated netlist using two randomly selected keys, and build a miter circuit that evaluates to 1 only when the output of two circuits is different. The miter circuit is then fed to a SAT solver looking for such inputs. The SAT returns with minimum of two inputs for which the outputs are different. These input patterns are tested using an activated IC (golden IC) validating the correct output. Then a bypass circuit is constructed using a comparator that is stitched to the primary output of the netlist which is unlocked using the selected random key, to retrieve the correct functionality if that input pattern is applied. The Bypass attack works well when the SAT-hard solution is not mixed with traditional logic locking mechanism since its overhead increases very quickly as output corruption of logic locking increases. This observation motivated researchers to look at possibilities of approximate attacks to retrieve the key values associated to non SAT-hard obfuscation solutions that are mixed with SAT-hard solutions.



\subsection{AppSAT Attack}

So far, defences solution to mitigate the SAT attack, are based on the assumption that the attacker needs an exact attack on logic locking. However, Shamsi \emph{et al.} \cite{shamsi2017appsat} proposed a new attack (AppSAT), which relax this constraint. AppSAT shown in Algorithm \ref{AppSAT}, is an approximate attack on logic locking based on the SAT attack and random testing. The authors use \emph{probably-approximate-correct} (PAC) model for formulating approximate learning problem. The exact SAT attack continues to find DIPs until no more DIPs can be found. However, the AppSAT will be terminated in any early step in which the error falls below the certain limit. If this condition happens, the key value will be considered as an approximate key with specified error rate; otherwise, the random sampling that resulted in a disagreement will be added to a SAT formula as a new constraint. In AppSAT, heuristic methods for estimating the error is used for large functions, to avoid any computation complexity.


\subsection{Double-DIP Attack}

Double-DIP \cite{shen2017double} is another approximate attack, shown in Algorithm \ref{Double-DIP}. Double-DIP is an extension of SAT attack in which during each iteration, the discriminating input should eliminate at least two wrong keys. To illustrate its effectiveness, researchers used double-DIP to target SARLock$+$SSL, representing a compound of SAT-hard and high output corruption obfuscation. When the double-DIP attack terminates, the key of the traditional logic locking (SSL) is guaranteed to be correct. As a result, the compound logic locking will be reduced to a single SAT attack resilient technique, that could then be attacked using bypass attack. 


\begin{algorithm}
\caption{AppSAT Attack Algorithm \cite{shamsi2017appsat}}
\label{AppSAT}
\begin{algorithmic}[1]
\scriptsize
\Function{AppSAT\_Attack}{Circuit~C$_{L}$, Circuit C$_{O}$}
    \State \emph{i} $\gets$ 0; \emph{F$_{0}$} $\gets$ C$_{L}$(X, K$_{1}$, Y$_{1}$) $\land$ C$_{L}$(X, K$_{2}$, Y$_{2}$);
    \While {\emph{SAT}(\emph{F$_{i}$} $\land$ (Y$_{1}$ $\neq$ Y$_{2}$))}
        \State X$_d$[i] $\gets$ sat\_assignment (F$_{i} \land $(Y$_1 \neq $Y$_2$)); Y$_d$[i] $\gets$ C$_{O}$(X$_d$[i]);
        \State \emph{F$_{i+1}$} $\gets$ \emph{F$_{i}$} $\land$ C$_{L}$(X$_d$[i], K$_{1}$, Y$_d$[i]) $\land$ C$_{L}$(X$_d$[i], K$_{2}$, Y$_d$[i]); \emph{i} $\gets$ \emph{i+1};
        
        \State every $n$ rounds do
        \ForEach{($x \in $ Random Patterns)}
            \If {C$_{L}$(X, K$_{1}$, Y) $\neq$ C$_{O}$(X)}
               \State \emph{FailedPatterns} $\gets$ \emph{FailedPatterns} + 1;        \State \emph{F$_{i+1}$} $\gets$ \emph{F$_{i+1}$} $\land$ (C$_{L}$(X, K$_{1}$, Y) = C$_{O}$(X)); \emph{i} $\gets$ \emph{i+1};
            \EndIf   
        \EndFor
        \If {error < ErrorThreshold}
            \State return K$_{1}$ as an approximate key
        \EndIf
    \EndWhile
    \State \emph{$K^*$}  $\gets$ sat\_assignment$_{K_1}$(\emph{F$_{i}$});
\EndFunction
\end{algorithmic}
\end{algorithm}

\begin{algorithm}
\caption{Double-DIP Attack Algorithm \cite{shen2017double}}
\label{Double-DIP}
\begin{algorithmic}[1]
\scriptsize
\Function{DoubleDIP\_Attack}{Circuit~C$_{L}$, Circuit C$_{O}$}
    \State \emph{i} $\gets$ 0; \emph{F$_{0}$} $\gets$ C$_{L}$(X, K$_{1}$, Y$_{1}$) $\land$ C$_{L}$(X, K$_{2}$, Y$_{2}$) $\land$ C$_{L}$(X, K$_{3}$, Y$_{1}$) $\land$ C$_{L}$(X, K$_{4}$, Y$_{2}$) ;
    \While {\emph{SAT}(\emph{F$_{i}$} $\land$ (Y$_{1}$ $\neq$ Y$_{2}$)) $\land$ (K$_{1}$ $\neq$ K$_{3}$)) $\land$ (K$_{2}$ $\neq$ K$_{4}$))}
        \State X$_d$[i] $\gets$ sat\_assignment (F$_{i} \land $(Y$_1 \neq $Y$_2$)) $\land$ (K$_{1}$ $\neq$ K$_{3}$)) $\land$ (K$_{2}$ $\neq$ K$_{4}$)); \State Y$_d$[i] $\gets$ C$_{O}$(X$_d$[i]);
        \State \emph{F$_{i+1}$} $\gets$ \emph{F$_{i}$} $\bigwedge_{j=1}^{4}$ C$_{L}$(X$_d$[i], K$_{j}$, Y$_d$[i]); \emph{i} $\gets$ \emph{i+1};
    \EndWhile
    \State \emph{$K^*$}  $\gets$ sat\_assignment$_{K_1}$(\emph{F$_{i}$});
\EndFunction
\end{algorithmic}
\end{algorithm}


\subsection{Bit-Flipping Attack}

The Bit-flipping attack \cite{shen2018sat} is yet another attack against compound logic locking schemes in which a SAT-hard solution such as SARLock is combined with a traditional logic locking to guarantee both high error rates and resilience to the SAT-based attack. In Bit-flipping attack, the keys are first separated into two groups ($k_1$ and $k_2$) by counting DIPs for two keys with hamming distance equal to one. The attack is motivated from the observation that in traditional logic locking, wrong key causes substantial wrong input-output pattern. However, the error rate of bit-flipping function is usually very small. As shown in Algorithm \ref{Bit-flipping}, after separation of keys, this attack fixes SAT-resilient keys, $k_2$, as a random number, and uses a SAT solver to find the correct key values for $k_1$. After finding $k_1$, the bypass attack is applied to retrieve the original circuit. 



\begin{algorithm}
\caption{Bit-flipping Attack Algorithm \cite{shen2018sat}}
\label{Bit-flipping}
\begin{algorithmic}[1]
\scriptsize
\Function{BitFlipping\_Attack}{Circuit~C$_{L}$, Circuit C$_{O}$}
    \ForEach {$j$ < \emph{Fixed-iteration}}
        \State K$_{A}$ $\gets$ a random key;
        \ForEach {bit b $\in$ K$_{A}$}
            \State K$_{B}$ $\gets$ K$_{A}$ while bit b flipped;
            \State \emph{i} $\gets$ 0; \emph{F$_{0}$} $\gets$ C$_{L}$(X, K$_{A}$, Y$_{A}$) $\land$ C$_{L}$(X, K$_{B}$, Y$_{B}$); 
            \While {\emph{SAT}(\emph{F$_{i}$} $\land$ (Y$_{A}$ $\neq$ Y$_{B}$))} 
                \State X$_d$[i] $\gets$ sat\_assignment (F$_{i} \land $(Y$_A \neq $Y$_B$)); 
                \State \emph{F$_{i+1}$} $\gets$ \emph{F$_{i}$} $\land$ (X $\neq$ X$_d$[i]); \emph{i} $\gets$ \emph{i+1};
                \If{i > Threshold}
                    \State b is in K$_{1}$, 
                    \State \textbf{break};
                \EndIf
            \EndWhile
        \EndFor
        $j$ $\gets$ $j$ + 1;
    \EndFor 
    \State K$_{2}$ $\gets$ all key bits $/$  K$_{1}$; \Comment{Seperation is Done. Then, fix K$_{2}$ as a random number.}
    \State K$_{1}$ $\gets$ SAT\_ATTACK (C$_{L}$, C$_{O}$); \Comment{Find Traditional Keys using SAT.}
    \State C$^*_{L}$ $\gets$ update\_netlist(C$_{L}$ | K$_{1}$)
    \State \Return(BYPASS\_ATTACK(C$^*_{L}$);
\EndFunction
\end{algorithmic}
\end{algorithm} 

\subsection{AppSAT Guided Removal Attack}

AppSAT Guided Removal (AGR) attack targets compound logic locking, particularly Anti-SAT + traditional logic locking \cite{yasin2017removal}. This attack integrates AppSAT with a simple structural analysis of the locked netlist (a post-processing steps). Unlike AppSAT, the AGR attack recovers the correct key. In this attack, first the AppSAT is used to find the key of the traditional obfuscation scheme (used as a part of compound lock). Then, AGR targets the remaining key bits belong to the SAT-resilient logic locking, such as Anti-SAT block, through a simple structural analysis. As shown in Algorithm \ref{AGR}, in the post-processing steps, AGR finds the gate ($G$) at which most of the Anti-SAT key bits converge. AGR finds $G$ by tracing the transitive fanout of the Anti-SAT key inputs, where all the Anti-SAT key bits converge. The ratio of key bits converging at each of the inputs of the gate $G$, are close to $0.5$, which is shown as the \emph{selected property} in line 7 of Algorithm \ref{AGR}. AGR identifies the candidates for gate $G$ by checking this property for all gates in the circuit, and then sort these candidate based on the number of key inputs that converge at a gate and pick the gate $G$ from all candidates, which has the most number of key inputs converge to that gate. Then the attacker re-synthesize the design with the constant value for the output of $G$ gate and retrieving the correct functionality. 

\begin{algorithm}
\caption{AGR Attack Algorithm \cite{yasin2017removal}}
\label{AGR}
\begin{algorithmic}[1]
\scriptsize
\Function{AGR\_Attack}{Circuit~C$_{L}$, Circuit C$_{O}$}
    
    \State $\#$\emph{Cand} $\gets$ num\_gates(C$_{L}$)
    \While {($\#$\emph{Cand} > 1 and $!$\emph{Timeout})} 
        \State AppSAT\_Attack();   \Comment{4 times}
        \State \emph{Candidates} $\gets$ $\{\}$;
        \ForEach{\emph{gate} $\in$ C$_{L}$}
            \If{\emph{gate}$_{i}$ has the \emph{selected property}}
                \State \emph{Candidates} $\gets$ \emph{Candidates} + 1;
            \EndIf
        \EndFor
    \EndWhile
    \State $G$ $\gets$ Find\_Max\_key\_count(\emph{Candidates});
    \State C$_{Lock}$ $\gets$ remove\_\textbf{TFI}(C$_{L}$, $G$); \Comment{remove Transitive FanIn of the gate $G$}
    \State \Return C$_{Lock}$; \Comment{C$_{Lock}$: C$_{L}$ after removing Anti\_SAT block}
    
\EndFunction
\end{algorithmic}
\end{algorithm}

\subsection{Sensitization Guided SAT Attack}
 
While the one-point flipping circuit in Anti-SAT and SARLock are completely decoupled from the original netlist, Li \emph{et al.} \cite{li2017provably} proposed the AND-tree Insertion (ATI), as a SAT-resilient logic locking, which embeds AND trees inside the original netlist. It not only makes all aforementioned attack less effective, it also decreases the implementation overhead. Additionally, the input of AND-tree are camouflaged by inserting INV/BUF camouflaged gates, which can be replaced with the XOR/XNOR gates in order to lock the AND-tree. However, this defense was broken by a new attack that was coined as Sensitization Guided SAT (SGS) attack \cite{yasin2017removal}. The SGS attack is carried out in two stages: (1) \emph{sensitization} that exploits bias in input patterns which allows an attacker to apply only a subset of DIPs, i.e., those that bring unique values to the AND-tree inputs. (2) \emph{SAT attack} using the patterns discovered in the first stage.

\subsection{Functional Analysis Attack}

\begin{figure}[t]
\begin{minipage}{0.33\linewidth}
\centering
\includegraphics[width=\linewidth]{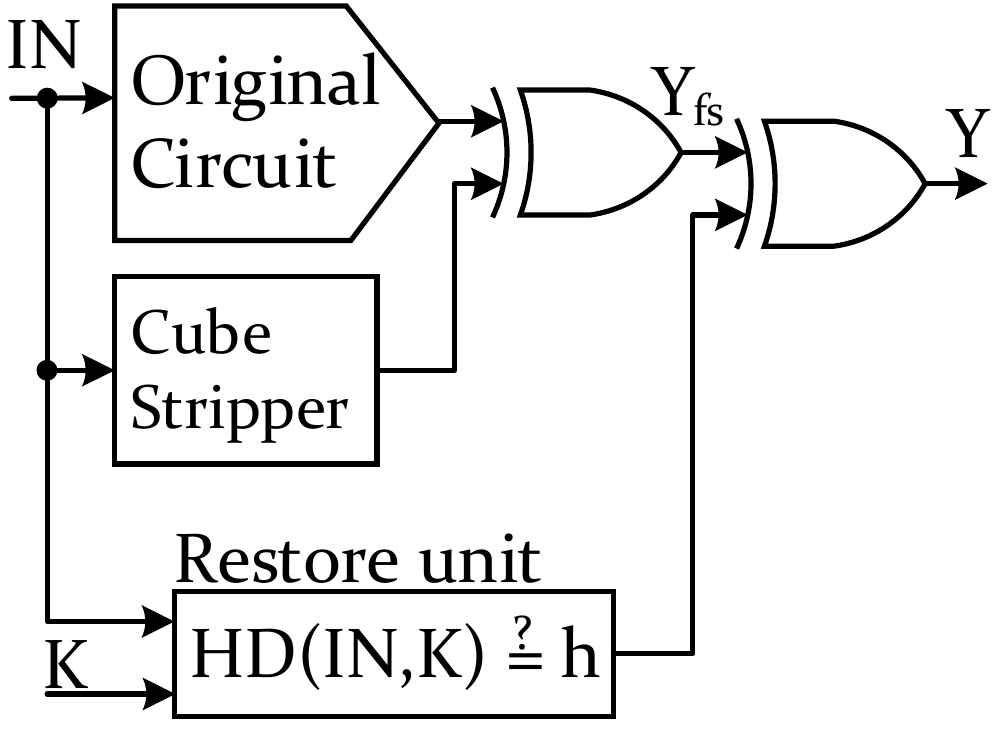}
\end{minipage}
\hspace{35pt}
\begin{minipage}{0.47\linewidth}
\scriptsize
\centering
\setlength\tabcolsep{2.5pt} 
\begin{tabular}{@{} l *{10}c @{}}
\toprule
$Y_{fs}$ & IN & k=0 & k=1 & k=2 & k=3 & k=4 & k=5 & k=6 & k=7 \\
\midrule
   \cmark    &  0 & \cmark & \cmark & \cmark & \cmark & \textcolor{red}{\xmark} & \cmark & \cmark & \cmark \\
   \cmark    &  1 & \textcolor{red}{\xmark} & \cmark & \cmark & \cmark & \cmark & \cmark & \cmark & \cmark \\
   \cmark    &  2 & \cmark & \cmark & \textcolor{red}{\xmark} & \cmark & \cmark & \cmark & \cmark & \cmark \\
   \cmark    &  3 & \cmark & \cmark & \cmark & \cmark & \cmark & \textcolor{red}{\xmark} & \cmark & \cmark \\
   \textcolor{red}{\xmark}   &  4 & \textcolor{red}{\xmark} & \cmark & \textcolor{red}{\xmark} & \textcolor{red}{\xmark} & \textcolor{red}{\xmark} & \textcolor{red}{\xmark} & \textcolor{red}{\xmark} & \textcolor{red}{\xmark} \\
   \cmark    &  5 & \cmark & \cmark & \cmark & \cmark & \cmark & \cmark & \textcolor{red}{\xmark} & \cmark \\
   \cmark    &  6 & \cmark & \cmark & \cmark & \cmark & \cmark & \cmark & \cmark & \textcolor{red}{\xmark}\\
   \cmark    &  7 & \cmark & \cmark & \cmark & \textcolor{red}{\xmark} & \cmark & \cmark & \cmark & \cmark \\ \bottomrule
\end{tabular}
\end{minipage}\hfill
\caption{SFLL-HD while $h = 0$.}
\label{sfll}
\end{figure}

Aiming to provide a defense that resists all previously formulated attacks led to the introduction of Stripped-Functionality Logic Locking (SFLL) \cite{yasin2017provably}. In SFLL the original circuit is modified for at-least one input pattern (cube) using a \emph{cube stripping unit}, demonstrated in Fig. \ref{sfll}. As shown, $Y_{fs}$ is the output of the stripped circuit, in which the output corresponding to at-least one input pattern is flipped. The restore unit not only generates the flip signal for one input pattern per each wrong key, it also restores the stripped output, (e.g. $IN = 4$ in Fig. \ref{sfll}) to recover the correct functionality on $Y$. Note that applying removal attack on restore unit recovers $Y_{fs}$, which is not the correct functionality. In addition, SFLL-HD is able to protect ${k}\choose{h}$ input patterns that are of Hamming Distance (HD) $h$ from the $k$-bit secret key, and accordingly uses Hamming Distance checker as a restore unit (e.g. $h=0$ in Fig. \ref{sfll} is also called TTLock \cite{yasin2017lock}). 

Although SFLL was resilient against all previously formulated attacks, it was exploited using a newly formulated attack, called Functional Analysis on Logic Locking (FALL) attack \cite{subramanyan1functional}. In this attack model, the adversary is assumed to be a malicious foundry that knows the locking algorithm and its parameters, e.g. $h$ in SFLL-HD. A FALL attack is carried out in three main stages and relies on structural and functional analyses to determine potential key values of a locked circuit. First, FALL attack tries to find all nodes which are the results of comparing an input value with a key input. It is done by a comparator identification. Such nodes ({$nodes_{RU}$}), which contains these particular comparators, are very likely to be part of the functionality restoration unit. The set of all inputs that appear in these comparators, should be in the fan-in cone of the cube stripping unit. Then, it finds a set of all gates whose fan-in-cone is identical to the members of {$nodes_{RU}$}. This set of gates must contain the output of the cube stripping unit. Second, the attacker applies functional analysis on the candidate nodes suggested by and collected from the first stage to identify suspected key values. Broadly speaking, the attacker uses functional properties of the cube stripping function used in SFLL, to determines the values of the keys. Based on the author's view, this function has three specific properties. So, they have proposed three attacks algorithms on SFLL, which exploit unateness and Hamming distance properties of the cube stripping functions. The input of these algorithm is circuit node $c$, that computed from the first stage, and the algorithm checks if $c$ behaves as a Hamming distance calculator in the cube stripping unit of SFLL-HD. If the attack is successful, the return value is the protected cube. Third, they have proposed a SAT-based key confirmation algorithm using a list of suspected key values and I/O oracle access, that verifies whether one of the suspected key values computed from the second stage, is correct. 

\subsection{CycSAT Attack}

Considering the strength of all previously formulated attacks, some of the researchers started seeking solutions that fundamentally violated the assumptions of these attacks with respect to the nature of locked circuits. One of such attempts was the introduction of cyclic logic locking \cite{shamsi2017cyclic}\cite{SRCLock}, was first proposed in \cite{shamsi2017cyclic}. In this obfuscation technique, as shown in Algorithm \ref{CycSAT_algoritm}, each deliberately established cycle is designed to have more than one way to open. The requirement for having more than one way to open each cycle assures that even if the original netlist has no cycle by itself,  the cycles remains irreducible by means of structural analysis. The cyclic obfuscation resulted in an obfuscation with high level of output corruption, while it was able to break the SAT attack either by 1) trapping it in an infinite loop, or 2) forcing it to exit with a wrong key depending on weather the introduced cycles make the circuit stateful or oscillating.   

The promise of secure cyclic obfuscation was shortly after broken by CycSAT attack \cite{zhou2017cycsat}.  In CycSAT, the key combinations that result in formation of cycles are found in a pre-processing step. These conditions are then translated into problem augmenting CNF formulas, denoted as cycle avoidance clauses, satisfaction of which guarantee no cycle in the netlist.  The cycle avoidance clauses are then added to the original SAT circuit CNF and the SAT attack is executed. The validity of this attack, however, was challenged in \cite{SRCLock}, as researchers illustrated that the pre-processing time for CycSAT attack is linearly dependent on the number of cycles in the netlist. Hence, by building an exponential relation between the number of feedback, and the number of cycles in the design, the pre-processing step of CycSAT will face exponential runtime. 

\subsection{Behavioral SAT (BeSAT) Attack}


Inability to analyze all cycles in the prepossessing step of CycSAT results in missing cycles in the pre-processing step of CycSAT, leading to building a statefull or oscillating circuit, trapping the SAT stage of the CycSAT attack. BeSAT \cite{shen2019besat} remedies this shortcoming by augmenting the CycSAT attack with a run-time behavioral analysis. As shown in Algorithm \ref{besat_algorithm}, by performing behavioral analysis at each SAT iteration, BeSAT detects repeated DIPs when the SAT is trapped in an infinite loop. Also, when SAT cannot find any new DIP, a ternary-based SAT is used to verify the returned key as a correct one, preventing the SAT from exiting with an invalid key.

\begin{algorithm}
\caption{CycSAT Attack on Cyclic Locked Circuits \cite{shamsi2017cyclic} \label{CycSAT_algoritm}}
\begin{algorithmic}[1]
\scriptsize

\Function{CycSAT\_Attack}{Circuit~C$_{L}$, Circuit C$_{O}$}
    \State $W = (w_0, w_1, ...w_m)$ $\gets$ FindFeedback(C$_{L}$);
    \ForEach{$(w_i \in W)$}
        \State $F(w_i, w'_i)$ $\gets$ no\_ structural\_ path($w_i$);
    \EndFor
    \State \emph{i} $\gets$ 0; $NC$(K)=$\wedge^m_{i=0}F(w_i,w'_i)$
    \State C$^*_{L}$(X, K, Y) $\gets$ C$_{L}$(X, K, Y) $\wedge$ $NC$(K); \emph{F$_{0}$} $\gets$ C$^*_{L}$(X, K$_{1}$, Y$_{1}$) $\land$ C$^*_{L}$(X, K$_{2}$, Y$_{2}$);
    \While {\emph{SAT}(\emph{F$_{i}$} $\land$ (Y$_{1}$ $\neq$ Y$_{2}$))} 
        \State X$_d$[i] $\gets$ sat\_assignment (F$_{i} \land $(Y$_1 \neq $Y$_2$)); Y$_d$[i] $\gets$ C$_{O}$(X$_d$[i]);
        \State \emph{F$_{i+1}$} $\gets$ \emph{F$_{i}$} $\land$ C$_{L}$(X$_d$[i], K$_{1}$, Y$_d$[i]) $\land$ C$_{L}$(X$_d$[i], K$_{2}$, Y$_d$[i]); \emph{i} $\gets$ \emph{i+1};
    \EndWhile
    \State \emph{$K^*$}  $\gets$ sat\_assignment$_{K_1}$(\emph{F$_{i}$});

\EndFunction

\end{algorithmic}
\end{algorithm}

\begin{algorithm}
\caption{BeSAT Attack on Cyclic Locked Circuits \cite{shen2019besat} \label{besat_algorithm}}
\begin{algorithmic}[1]
\scriptsize
\Function{BeSAT\_Attack}{Circuit~C$_{L}$, Circuit C$_{O}$}
    \State $W = (w_0, w_1, ...w_m)$ $\gets$ FindFeedback(C$_{L}$);
    \ForEach{$(w_i \in W)$}
        \State $F(w_i, w'_i)$ $\gets$ no\_ structural\_ path($w_i$);
    \EndFor
    \State \emph{i} $\gets$ 0; $NC$(K)=$\wedge^m_{i=0}F(w_i,w'_i)$
    \State C$^*_{L}$(X, K, Y) $\gets$ C$_{L}$(X, K, Y) $\wedge$ $NC$(K); \emph{F$_{0}$} $\gets$ C$^*_{L}$(X, K$_{1}$, Y$_{1}$) $\land$ C$^*_{L}$(X, K$_{2}$, Y$_{2}$);
    \While {\emph{SAT}(\emph{F$_{i}$} $\land$ (Y$_{1}$ $\neq$ Y$_{2}$))} 
        \State X$_d$[i] $\gets$ sat\_assignment (F$_{i} \land $(Y$_1 \neq $Y$_2$)); Y$_d$[i] $\gets$ C$_{O}$(X$_d$[i]);
        \State \emph{F$_{i+1}$} $\gets$ \emph{F$_{i}$} $\land$ C$_{L}$(X$_d$[i], K$_{1}$, Y$_d$[i]) $\land$ C$_{L}$(X$_d$[i], K$_{2}$, Y$_d$[i]);
        \If{(X$_d$[i] in DIP) \textbf{and} (C$_{L}$(X$_d$[i], K$_{1}$) $\neq$ Y$_d$[i]))}
            \State \emph{F$_{i+1}$} $\gets$ \emph{F$_{i+1}$} $\land$ (K$_{1}$ $\neq$ \^{K$_{1}$}) $\land$ (K$_{2}$ $\neq$ \^{K$_{1}$});
        \ElsIf{(X$_d$[i] in DIP) \textbf{and} (C$_{L}$(X$_d$[i], K$_{2}$) $\neq$ Y$_d$[i])}
                \State \emph{F$_{i+1}$} $\gets$ \emph{F$_{i+1}$} $\land$ (K$_{1}$ $\neq$ \^{K$_{2}$}) $\land$ (K$_{2}$ $\neq$ \^{K$_{2}$});
        \EndIf
        \State \emph{i} $\gets$ \emph{i+1};
    \EndWhile
    \While {\emph{SAT}$_{K_1}$(F$_{i}$)} \Comment{Correct Key: \^{K$_{c}$}}
        \If{Ternary\_SAT(F$_{i}$, K$_{c}$)}
            \State F$_{i}$ $\gets$ F$_{i}$ $\land$  (K$_{1}$ $\neq$ \^{K$_{c}$})
        \Else
            \State \emph{$K^*$}  $\gets$ \^{K$_{c}$}; \textbf{break};
        \EndIf
    \EndWhile
\EndFunction

\end{algorithmic}
\end{algorithm}

\section{Stage 4: SMT Attack} \label{stage4}

\begin{figure}[t]
\centering
\includegraphics[width = 350pt]{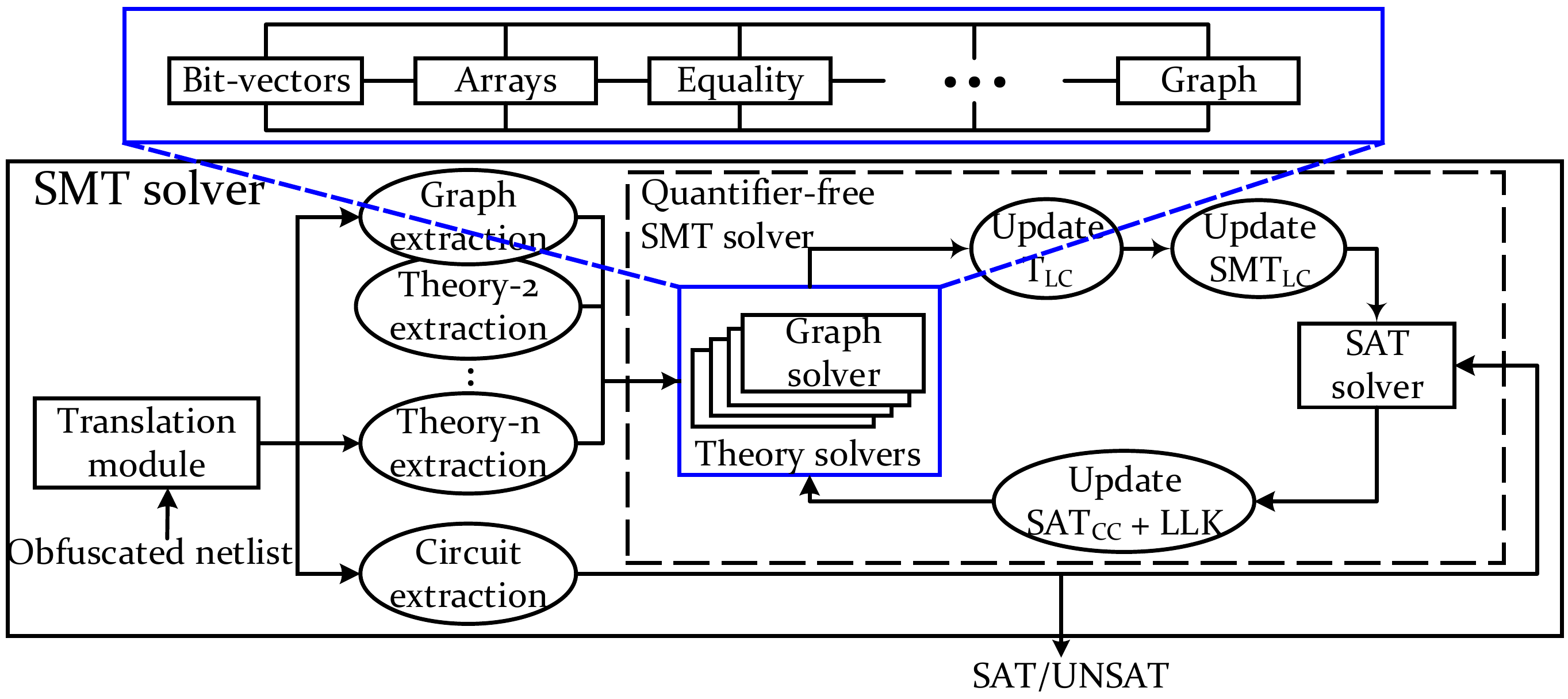}
\caption{Overall Architecture of SMT Attack for Circuit Deobfuscation \cite{azar2019smt}.}
\label{TopArch}
\end{figure}

As discussed previously, many of the attacks proposed at post-SAT attack stage were formulated by adding a pre-processing step to the original SAT attack, and/or extending the SAT attack to co-process and check additional features in each iteration. In other terms, to break many of the post-SAT era obfuscation techniques, attackers relied on compound attacks by combining SAT solvers by pre-processors (e.g. in CycSAT) and co-processors (e.g. in BeSAT) to extend its modeling reach. Motivated by this trend, the need for having pre- co- and post- processors along with a SAT solver in an attack was realized and addressed in \cite{azar2019smt} and a new and extremely powerful attack, coined as Satisfiability Module Theory (SMT) attack was introduced. The strength of SMT attack, as the superset of SAT attack, comes from its ability to combine SAT and Theory solvers. As shown in Fig. \ref{TopArch}, The SMT attack could be invoked with any number and combination of theory solvers, and a SAT solver, which allow the attacker to express constraints that are difficult or even impossible to express using CNF, including timing, delay, power, arithmetic, graph and many other first-order theories in general. To showcase the modeling capability of SMT attack, the authors used the SMT attack 1) to break a new breed of obfuscation that relied on locking the delay information in netlist (by generating setup and hold violations), 2) to formulate an accelerated attack (to reduce the attack time) with means of approximate exit (if trapped with SAT hard solutions).

In pursuit of obfuscation schemes that could not be attacked by SAT motivated attackers, some researchers tried to extend the locking mechanism to aspects of a circuit's function that cannot be translated to CNF. For example, Xie \emph{et al.} proposed a timing obfuscation scheme, denoted as delay logic locking (DLL), in \cite{xie2017delay}. The Goal of DLL obfuscation scheme is introducing setup and hold violation if the correct key is not applied. In this case, the obfuscation attempts to change both logical and behavioral (timing) properties. A functionally-correct but timing-incorrect key will result in timing violations, leading to circuit malfunctions. Considering that timing is not translatable to CNF, the SAT solver remains oblivious to the keys used for timing obfuscation.  Authors in \cite{azar2019smt}, however, illustrated that the SMT attack could easily deploy a graph theory solver, provide timing constraints to the theory solver (in terms of required min and max delay to meet the hold and setup time), and use the theory solver in parallel with the internal SAT solver to break both logic and delay obfuscation. They additionally show that the theory solver could be initiated as a pre-processor (Eager SMT approach) or as a co-processor (Lazy SMT approach) to break the same problem, showcasing the strength of SMT attack. The \emph{lazy} mode of this attack is illustrated in Algorithm \ref{smt_lazy}. Although at about the same time Chakraborty proposed TimingSAT to attack the DLL \cite{chakraborty2018timingsat}, similar to many prior SAT-based attack, it was by deploying a pre-processor for analysis of graph timing, and generating helper clauses for the subsequent call to the SAT attack.

\begin{algorithm}
\caption{SMT Attack on DLL (Lazy Approach) \cite{azar2019smt}}
\label{smt_lazy}
\begin{algorithmic}[1]
\scriptsize
\Function{SMTLazy\_Attack}{Circuit~C$_{L}$, Circuit C$_{O}$}
    \State C$^*_{L}$ $\gets$ toBOOLEAN(C$_{L}$); \Comment{Replace TDK with Buffer}
    \State \emph{i} $\gets$ 0; \emph{F} $\gets$ C$^*_{L}$(X, K$_{1}$, Y$_{1}$) $\land$ C$^*_{L}$(X, K$_{2}$, Y$_{2}$); 
    \State G$^*_{L}$ $\gets$ toGRAPH(C$_{L}$); \Comment{Wires = Edges, Gates = Vertices}
    \State \emph{F$_T$} $\gets$ GenTCE(G$^*_{L}$) \Comment{Theory Learned Clauses}
    \State \emph{F$_{SMT}$} $\gets$ \emph{F} $\land$ \emph{F$_T$}; \Comment{SMT Clauses}
    \While {\emph{SMT}(\emph{F$_{SMT}$})} \Comment{X$_d$[i], K$_{1}$, K$_{2}$, CC}
        \State Y$_d$[i] $\gets$ C$_{O}$(X$_d$[i]); \emph{F} $\gets$ \emph{F} $\land$ C$^*_{L}$(X$_d$[i], K$_{1}$, Y$_d$[i]) $\land$ C$^*_{L}$(X$_d$[i], K$_{2}$, Y$_d$[i]);
        \State \emph{F$_{SMT}$} $\gets$ \emph{F} $\land$ CC; \emph{i} $\gets$ \emph{i+1};
    \EndWhile
    \State \emph{$K^*$}  $\gets$ smt\_assignment$_{K_1}$(\emph{F$_{SMT}$});

\Statex \hrulefill
\setcounter{ALG@line}{0}

    \Function{GenTCE}{Graph G$^*_{L}$}
        \State \emph{Inputs} $\gets$ find\_start\_points(G$^*_{L}$); \emph{Outputs} $\gets$ find\_end\_points(G$^*_{L}$); \emph{T$_{CE}(K)$} $\gets$ [];
        \ForEach{(\emph{(Sp, Ep)} $\in$ (\emph{Inputs, outputs})}
        	\State Upper(Sp,Ep)(K) $\gets$ !(\textbf{distance\_leq}(Sp, Ep, t$_{cd}$)); \Comment{Hold Violation}
        	\State Lower(Sp,Ep)(K) $\gets$ \textbf{distance\_leq}(Sp, Ep, t$_p$); \Comment{Setup Violation}
    		\State \emph{Range(Sp,Ep)(K)} $\gets$ Lower(Sp,Ep)(K) $\land $ Upper(Sp,Ep)(K);	
    		\State \emph{T$_{CE}(K)$} $\gets$ \emph{T$_{CE}(K)$} $\cup$ \emph{Range(Sp,Ep)(K)};
        \EndFor
        \State \Return \emph{T$_{CE}(K)$}
    \EndFunction

\EndFunction
\end{algorithmic}
\end{algorithm}


The ability of SMT solver to instantiate and integrate different theory solver makes it a suitable attack platform for modeling and formulating very strong attacks. As an example of the strength of SMT attack, the authors in  \cite{azar2019smt} formulated and presented an accelerated SMT attack with ability of detecting the presence of SAT-hard obfuscation and switching to an accelerated approximate attack. As shown in Algorithm \ref{smt_accelerated}, this was done by invoking a \emph{BitVector} theory solver to constrain the SMT solver for finding keys that result in highest output corruption first. This could be done by constraining the required HD between the output of double circuit when two different keys for the same discriminating input is being tested. The required HD starts from a large value, and every time that the SMT solver return UNSAT, the constraint is relaxed until HD of 1 is reached. This leads to the guaranteed discovery of keys for traditional logic locking first. After $N$ tries ($Rep$ in Algorithm \ref{smt_accelerated}) for HD of 1, the SMT attack exits, notes that there exist a SAT-hard obfuscation, which now could be addressed by the Bypass attack. 

\begin{algorithm}
\caption{Accelerated SMT Attack on Compound Locking \cite{azar2019smt}}
\label{smt_accelerated}
\begin{algorithmic}[1]
\scriptsize
\Function{AccSMT\_Attack}{Circuit~C$_{L}$, Circuit C$_{O}$}
    \State \emph{i} $\gets$ 0; \emph{HD$_{h}$} $\gets$ \emph{sizeof}(output); \emph{HD$_{l}$} $\gets$ \emph{HD$_{h}$} - 1;
    \State \emph{TimeOut} $\gets$ 20; \emph{Rep} $\gets$ 20; \emph{HD$_R$} $\gets$ 1; \emph{R$_{cnt}$} $\gets$ 0;
    \State C$^*_{L}$ $\gets$ toBOOLEAN(C$_{L}$); \Comment{Everything is Boolean.}
    \State \emph{F} $\gets$ C$^*_{L}$(X, K$_{1}$, Y$_{1}$) $\land$ C$^*_{L}$(X, K$_{2}$, Y$_{2}$);
    \State BV$^*_{L}$ $\gets$ toBITVECTOR(C$_{L}$); \Comment{Define BITVECTOR on output.}
    \State BVs$^*_{L}$(X, K$_{1}$, K$_{2}$) $\gets$ ONEs(BV$^*_{L}$(X, K$_{1}$) $\oplus$ BV$^*_{L}$(X, K$_{2}$));
    \State \emph{F$_T$} $\gets$ (BVs$^*_{L}$(X, K$_{1}$, K$_{2}$) $\geqslant$ \emph{HD$_{l}$}) $\land$ (BVs$^*_{L}$(X, K$_{1}$, K$_{2}$) $\leqslant$ \emph{HD$_{h}$});
    \State \emph{F$_{SMT}$} $\gets$ \emph{F} $\land$ \emph{F$_T$}; \Comment{SMT Clauses}
    \While{\emph{HD$_{l}$} $\geqslant$ 1}
        \While {\emph{SMT}(\emph{F$_{SMT}$} | \emph{TimeOut})} \Comment{X$_d$[i], K$_{1}$, K$_{2}$, CC}
            \State Y$_d$[i] $\gets$ C$_{O}$(X$_d$[i]);
            \State \emph{F} $\gets$ \emph{F} $\land$ C$^*_{L}$(X$_d$[i], K$_{1}$, Y$_d$[i]) $\land$ C$^*_{L}$(X$_d$[i], K$_{2}$, Y$_d$[i]); \emph{F$_{SMT}$} $\gets$ \emph{F} $\land$ CC;
            \If{\emph{HD$_{l}$} $\leqslant$ \emph{HD$_R$}}
                \If{\emph{R$_{cnt}$} == \emph{Rep}}
                    \State \textbf{break};
                \EndIf
                \State \emph{R$_{cnt}$}++;
            \EndIf
        \EndWhile
        \emph{HD$_{l}$}-\--;
    \EndWhile
    \State \emph{$K^*$}  $\gets$ smt\_assignment$_{K_1}$(\emph{F$_{SMT}$});
\EndFunction
\end{algorithmic}
\end{algorithm}

\section{Discussion \& Opportunities} \label{stage5}


Table \ref{compare} compares the effectiveness of the attacks discussed in this paper against most notable obfuscation schemes. As illustrated the combination of FALL, Bypass and SMT attack can break all existing solutions, pointing us to a need for a new direction for generating non-bypassable SMT hard obfuscation solutions. The dilemma is that SAT-hard solutions have extremely low output corruption, and are prone to Bypass, FALL, Removal and SPS attack. On the other hand, the traditional logic locking schemes have high output corruption, but could be easily broken with SAT/SMT attack. The compound logic locking solutions that combine the SAT-hard solutions for resistance against SAT and SMT attack, and traditional logic locking for resistance against Bypass, FALL, Removal and SPS attack are also prone to approximate SAT and SMT attacks. What is really desired, is a SMT-hard logic locking scheme with high degree of output corruption. As a step in this direction, few very recent research papers have focused on increasing the execution time of each SAT/SMT iteration rather than the total execution time \cite{kamali2019full}, \cite{shamsi2018cross}. The Full-Lock in \cite{kamali2019full} is argued that the strength of SAT/SMT solvers come from their \emph{Conflict-Driven Clause Learning} (CDCL) ability, which is resulted by recursively calling  \emph{Davis-Putnam-Logemann-Loveland} (DPLL) algorithm. Hence, the Full-Lock creates an obfuscation method that results in very deep recursive call trees. They argue that the SAT/SMT attack execution time can be expresses by formula \ref{runtime1}, in which $N$ denotes the number of iterations (DIPs) of the SAT/SMT attack, $T_{DPLL}(\Phi)$ is the execution time of recursive calls for DPLL algorithm on CNF $\Phi$, and $t$ is the execution time of the remaining book keeping code executed at each iteration.

\begin{equation} \label{runtime1} \scriptsize
T_{Attack}  = \sum_{i=1}^{N}T(i) = \sum_{i=1}^{N}(t + T_{DPLL}(\Phi)) = \sum_{i=1}^{N}\sum_{j=1}^{M}(T_{DPLL}^{Avg}) \simeq MN\times T_{DPLL}^{Avg}
\end{equation}

Authors argue that $M$ in formula \ref{runtime1} denotes the number of recursive DPLL calls. Accordingly, the execution time of SAT attack could also become unfeasible by building an exponential relation between the percentage gate inserted (area overhead) and M. The strong aspect of this alternative solution is that (1) the problems posed at each iteration of SAT/SMT attack is a SAT-hard problem, (2) the output corruption of this methods is significantly higher than obfuscating solution relying on increasing the $N$, (3) it is not susceptible to SPS, removal, bypass, approximate attack, to name a few. The hardness of SAT/SMT attack in the solution posed by Full-Lock cannot be assessed/formulated similar to that of SFLL. Moving towards this new direction for generating SAT-hard problems with high level of output corruption can be generalized more, where an obfuscation solution in this direction can engineer the number of recursive calls, pushing the number of recursive call to be an exponential function of added gates counts (area overhead).

\begin{table}[t]
\scriptsize
\centering
\caption{Comparison of proposed attacks/defenses.}
\label{compare}
\setlength\tabcolsep{1.3pt} 
\begin{tabular}{@{} l *{13}c @{}}
\hline 
\multirow{2}{*}{\backslashbox[40pt]{Attacks}{Defenses}}  & \textcolor{white}{.} & \textcolor{white}{.} & \textbf{\tiny{RLL}} & \textbf{\tiny{FLL}} & \textbf{\tiny{SLL}} & \textbf{\tiny{Anti-SAT}} & \textbf{\tiny{SARLock}} & \textbf{\tiny{Compound}} & \textbf{\tiny{SFLL}} & \textbf{\tiny{Cyclic}} & \textbf{\tiny{SRCLock}} & \textbf{\tiny{DLL}} \\
 & & & \cite{roy2010ending} & \cite{rajendran2015fault} & \cite{rajendran2012security} & \cite{xie2016antisat} & \cite{yasin2016sarlock} & \cite{yasin2016sarlock} & \cite{yasin2017provably} & \cite{shamsi2017cyclic} & \cite{SRCLock} & \cite{xie2017delay} \\

\midrule
   \textbf{Brute Force}        & & &   \xmark & \xmark & \xmark & \xmark & \xmark & \xmark & \xmark & \xmark & \xmark & \xmark \\
   
   \textbf{Sensitization} \cite{rajendran2012security}    & & &  \cmark &  \cmark & \xmark & \xmark & \xmark & \xmark & \xmark & \xmark & \xmark & \xmark \\
   
   \textbf{Hill-Climbing} \cite{plaza2015solving}    & & &  \cmark & \cmark & \xmark & \xmark & \xmark & \xmark & \xmark & \xmark & \xmark & \xmark \\
   
   \textbf{SAT} \cite{subramanyan2015evaluating}  & & &  \cmark & \cmark & \cmark & \xmark & \xmark & \xmark & \xmark & \xmark & \xmark & \xmark \\
   
   \textbf{SPS+Removal} \cite{yasin2017removal}\cite{Yasin2017sps}  & & &  \xmark & \xmark & \xmark & \cmark & \cmark & \xmark & \xmark & \xmark & \xmark & \xmark \\
   
   \textbf{Bypass} \cite{xu2017novel}  & & &  \xmark & \xmark & \xmark & \cmark & \cmark & \xmark & \xmark & \xmark & \xmark & \xmark \\
   
   \textbf{AppSAT} \cite{shamsi2017appsat} & & &  \cmark & \cmark & \cmark & \xmark & \xmark & \textbf{P} & \xmark & \xmark & \xmark & \xmark \\
   
   \textbf{Double-DIP} \cite{shen2017double}   & & &  \cmark & \cmark & \cmark & \xmark & \xmark & \textbf{P} & \xmark & \xmark & \xmark & \xmark \\ 
   
   \textbf{Bit-Flipping} \cite{shen2018sat}  & & &  \cmark & \cmark & \cmark & \cmark & \cmark & \cmark & \xmark & \xmark & \xmark & \xmark \\ 
   
   \textbf{AGR} \cite{yasin2017removal}   & & &  \cmark & \cmark & \cmark & \cmark & \cmark & \cmark & \xmark & \xmark & \xmark & \xmark \\ 
   
   \textbf{FALL} \cite{subramanyan1functional}  & & &  \xmark & \xmark & \xmark & \xmark & \xmark & \xmark & \cmark & \xmark & \xmark & \xmark \\ 
   
   \textbf{CycSAT} \cite{zhou2017cycsat}  & & &  \cmark & \cmark & \cmark & \xmark & \xmark & \xmark & \xmark & \cmark & \xmark & \xmark \\ 
   
   \textbf{BeSAT} \cite{shen2019besat}   & & &  \cmark & \cmark & \cmark & \xmark & \xmark & \xmark & \xmark & \cmark & \xmark & \xmark \\ 
   
   \textbf{TimingSAT} \cite{chakraborty2018timingsat}  & & &  \cmark & \cmark & \cmark & \xmark & \xmark & \xmark & \xmark & \xmark & \xmark & \cmark \\ 
   
   \textbf{SMT} \cite{azar2019smt}  & & &  \cmark & \cmark & \cmark & \xmark & \xmark & \textbf{P} & \xmark & \cmark & \cmark & \cmark \\ 
   \bottomrule
   
\multicolumn{13}{l}{\cmark: Attack Success, \xmark: Fail to Attack.} \\ \multicolumn{13}{l}{\textbf{P}: Only removes the key to the traditional locking in Compound Defense.}
\end{tabular}
\end{table}




\bibliographystyle{plain}
\bibliography{s-bibliography}

\end{document}